# STIMONT: A core ontology for multimedia stimuli description

Marko Horvat[1], Nikola Bogunović[1] and Krešimir Ćosić[1]

[1] University of Zagreb, Faculty of Electrical Engineering and Computing,
Unska 3, HR-10000 Zagreb, Croatia
E-Mail: Marko.Horvat2@fer.hr, Tel: +385-1-6129-565, Fax: +385-1-6129-705

**Abstract.** Affective multimedia documents such as images, sounds or videos elicit emotional responses in exposed human subjects. These stimuli are stored in affective multimedia databases and successfully used for a wide variety of research in psychology and neuroscience in areas related to attention and emotion processing. Although important all affective multimedia databases have numerous deficiencies which impair their applicability. These problems, which are brought forward in the paper, result in low recall and precision of multimedia stimuli retrieval which makes creating emotion elicitation procedures difficult and labor-intensive. To address these issues a new core ontology STIMONT is introduced. The STIMONT is written in OWL-DL formalism and extends W3C EmotionML format with an expressive and formal representation of affective concepts, high-level semantics, stimuli document metadata and the elicited physiology. The advantages of ontology in description of affective multimedia stimuli are demonstrated in a document retrieval experiment and compared against contemporary keyword-based querying methods. Also, a software tool Intelligent Stimulus Generator for retrieval of affective multimedia and construction of stimuli sequences is presented.

## 1 Introduction

Multimedia documents with annotated semantic and emotion content are stored in affective multimedia databases. Apart from digital objects these databases contain metadata about their high-level semantics and expected emotion that will be induced in a subject when exposed to a contained document. Two important features distinguish affective multimedia databases from other multimedia repositories: *i)* the purpose of the multimedia documents and *ii)* the emotion representation of the multimedia documents. Multimedia documents in affective multimedia databases are aimed at inducing or stimulating emotions in exposed subjects. As such they are usually referred to as stimuli. By being exposed to multimedia stimuli individuals' emotional states may be modulated [1, 2]. This spontaneous cognitive process can be utilized in various domains like Affective Computing and Human-Computer Interaction (HCI) but also in research of human emotions, attention and development of stress-related mental disorders. Affective multimedia databases are standardized which allows them to be used in a controllable and predictable manner and, subsequently, the emotion elicitation results can be measured, replicated and validated by different research teams.

This paper addresses multiple drawbacks of contemporary affective multimedia databases [3] and proposes an ontology-based approach for formal description of stimuli metadata which aims to optimize both the annotation and retrieval processes from these databases. Therefore, a new STIMulus ONTology (STIMONT) for a formal and comprehensive description of multimedia stimuli has been developed along with a software tool Intelligent Stimulus Generator for database searching, stimuli sequence construction and subject exposure. The ontology is validated and the retrieval results are compared with the existing methods for querying of affective multimedia databases. Motivation for this work was supported by an online survey on usage patterns of multimedia stimuli databases [4]. The survey indicated that domain experts unequivocally recognize the need for an intelligent stimuli retrieval application that would assist them in experimentation. Also, almost all experts agreed that such applications would be useful in their work.

The remainder of this paper is organized as follows; Section 2 gives an overview of the related work including most important contemporary stimuli databases, employed emotion theories and existing methods for affective multimedia annotation. In Section 2.1 a number of databases' deficiencies, which could be amended with an ontology-based formal knowledge presentation, are brought forward. Section 2.2 explains how ontology-based reasoning techniques can be utilized to achieve formal interpretation of high-level semantics in multimedia stimuli. Section 3 introduces the STIMONT model. All aspects of the proposed ontology are thoroughly described, in particular knowledge models of emotion, semantics, context and physiology domains. Usage of the ontology is demonstrated in several real-world examples. Implementation of the ontology is explained in Section 4, and an experimental validation using database querying and affective multimedia document retrieval is described in Section 4.1. Finally, Section 5 discusses various aspects of the proposed ontology and provides insight into future work.

## 2 Related work

This section gives a brief introduction to affective multimedia databases, multimedia emotion annotation models and the ontology-based methods for high-level metadata representation and document retrieval. The development of the ontology specifically tailored for stimuli description is directly motivated by the need to improve querying and retrieval from affective multimedia databases.

### 2.1 Affective multimedia databases

The International Affective Picture System (IAPS) [5] and the International Affective Digital Sounds system (IADS) [6] are two of the most cited databases in the area of affective stimulation. The latest versions of IAPS and IADS contain 1182 and 167 semantically and emotionally annotated stimuli, respectively. These databases cover a wide range of semantic categories characterized along the affective dimensions of pleasure, arousal and dominance. They were created with three goals in mind [7]:
1. Better experimental control of emotional stimuli;
2. Increasing the ability of cross-study comparisons of results;
3. Facilitating direct replication of undertaken studies.

The same standardization principles are shared among other similar affective multimedia databases. Apart from the IAPS and IADS the most frequently used and readily available affective multimedia databases are Geneva Affective PicturE Database (GAPED) [8], Nencki Affective Pictures System (NAPS) [9], Dataset for Emotion Analysis using eeg, Physiological and video signals (DEAP) [10], NimStim Face Stimulus Set [11], Pictures of Facial Affect (POFA) [12], Affective Norms for English Words (ANEW) [13], Affective Norms for English Texts (ANET) [14] and SentiWordNet [15]. Additional audio-visual affective multimedia databases with category or dimensional emotion annotations are listed here [16]. Facial expression databases are by far the most numerous modality among the affective multimedia databases and, although they are employed in emotion elicitation, facial expression databases are primarily used for face recognition and face detection. A more detailed overview of these databases is given in [17].

Two predominant theories used to describe emotion are the discrete category model and the dimensional model of affect (also sometimes called Circumplex model of affect [18] or PAD [19]). All affective multimedia databases have been characterized according to at least one of these models [20]. The dimensional theories of emotion propose that affective meaning can be well characterized by a small number of dimensions. Dimensions are chosen on their ability to statistically characterize subjective emotional ratings with the least number of dimensions possible [21]. These dimensions generally include one bipolar or two unipolar dimensions that represent positivity and negativity and have been labeled in various ways, such as valence or pleasure. Also usually included is a dimension that captures intensity, arousal, or energy level. In contrast to the dimensional theories, categorical theories claim that the dimensional models, particularly those using only two or three dimensions, do not accurately reflect the neural systems underlying emotional responses. Instead, supporters of these theories propose that there are a number of emotions that are universal across cultures and have an evolutionary and biological basis [22]. Which discrete emotions are included in these theories is a point of contention, as is the choice of which dimensions to include in the dimensional models. Most supporters of discrete emotion theories agree that at least the five primary emotions of happiness, sadness, anger, fear and disgust should be included.

Dimensional and categorical theories of affect can both effectively describe emotion in digital systems but are not mutually exclusive. Stimuli previously only characterized according to a single theory have also been characterized according to the complimentary emotion theory, as for example in IAPS [23], IADS [24] and ANEW [25]. Annotations according to both theories of affect are useful for several reasons, predominantly because they providing a more complete characterization of stimuli affect.

In terms of semantic content a single multimedia stimulus is described with a single tag from an unsupervised glossary. Semantic relations between different concepts are undefined and multiple different keywords could be used for description of the same concept. For example, a picture stimulus portraying an attack dog may be tagged as „dog", „attack", „attackdog", „attack_dog" etc. Synonyms like „canine" or „hound" would be interpreted as different concepts. Semantic lexicon has no semantic similarity measures and there are no criterions to estimate relatedness between concepts. In such multimedia stimuli databases it is impossible to establish that „dog" and „cat" are more closely related than „dog" and „Space Shuttle". This represents a huge defect in the stimuli retrieval process because a search query has to lexically match the database's keywords and no higher or more semantically meaningful interpretation of the query and annotating tags is attainable. The inadequate semantic descriptors result in three negative effects which impair stimuli retrieval: 1) low recall, 2) low precision and high recall or 3) vocabulary mismatch. Furthermore, affective multimedia databases contain only data about the stimuli themselves and do not describe semantics implied by the stimuli content. An example of the inadequacy of semantic descriptors in contemporary affective multimedia databases is illustrated in the figure below.

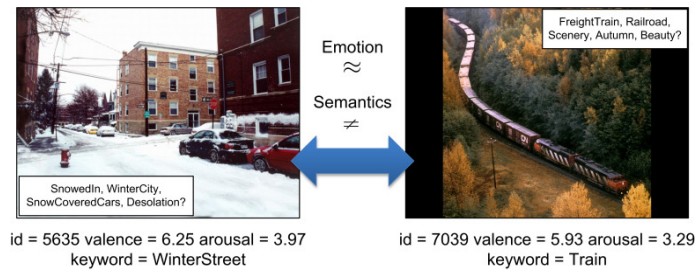

**Fig. 1.** An example of two IAPS pictures (5635.jpg, 7039.jpg) with similar emotion annotations but different and inadequately described high-level semantics. Stimuli with objects in context cannot be sufficiently well described with representation schemes in the existing stimuli databases.

Pictures $p_1$ and $p_2$ are taken from IAPS ($p_1$=5635.jpg, $p_2$=7039.jpg) and have PAD emotion values. Both pictures have closely similar neutral valence and sub neutral arousal $p_1: (valence, arousal) = (6.25, 3.97)$, $p_2: (valence, arousal) = (5.93, 3.29)$. As mentioned before these values have been methodically acquired and verified in a manual picture annotation experiment. Although both pictures have similar emotion values they have different semantics and would be appraised differently by subjects. High-level semantics of $p_1$ and $p_2$ – generally an object or a scene – are described with just one keyword "WinterStreet" and "Train", respectively. While this may be enough for representation of an isolated object, e.g. in a close-up picture with minimal context, this is clearly insufficient for rich meaning of objects within a scene even just for the most relevant and evident high-level semantics. For example, picture $p_1$ is tagged as "WinterStreet" but any of the following tags could substitute the original tag: "SnowedIn", "WinterCity", "SnowCoveredCars" or even "Desolation". Equally so, picture $p_2$ could be described as "FreightTrain", "Railroad", "Scenery", "Autumn" or "Beauty". Reliance on emotion values alone is not enough to discriminate stimuli but at the same time IAPS tagging system is inadequate in describing complete stimuli semantics.

In summary, available standardized affective multimedia databases offer a variety of quality audio-visual stimuli to researchers in the field. They were meticulously built and enable valid comparative measurements. But also they are mutually non-compliant, non-contiguous and structurally overly diversified. Their content is described loosely and informally. The databases are domain dependant and have arbitrarily structures.

### 2.2 Ontology-based representation of high-level stimuli content

By definition ontologies are a representation of a shared understanding about a specific domain and enable the derivation of implicit knowledge from the existing explicit knowledge and automated inference with reasoning engines [26]. Ontologies have been successfully applied for description of high-level image content, concept semantics, object labels and relationships defined in the upper levels of the image representation hierarchy [27]. This top-down approach in document representation and retrieval has three main benefits over the opposite approach (i.e. bottom-up) which relies on media features and other low level image descriptors. Firstly, database users prefer to articulate their search queries in a natural language, or in a constructed language similar to their preferred natural language, that are inherently capable for expression of a complex semantics. Secondly, the information one can infer from raw media information cannot be automatically transformed to high-level semantics that the stimuli convey. Thirdly, only rich high level full semantic representation of an image can express the full range of relationships, explicitly observable, implicitly inferable and the variety of purported connotations, actions and the broader context.

Except the W3C EmotionML format [28] much work has been done to enable describing information about emotions in multimedia, especially in the video. MPEG-7 multimedia standard, which is based on XML and can be expanded with additional tools, provides a method for describing emotions with its Affective Description Scheme [29]. Researchers have proposed new description tools that rely on MPEG-7 to provide a broader description of affective terms that can be used in video annotation [30]. Also, several multimedia ontologies which are potentially applicable in emotion description have been proposed in the last decade [31] including ontologies specially designed for high-level description of cognitive-emotional related concepts [32]. However, EmotionML currently offers the most sophisticated emotion annotation glossary and, since it is also XML compliant, can be used in MPEG-7 and other compatible standards.

The retrieval of multimedia stimuli is in many respects similar to retrieval of general multimedia, except for the added dimension of emotion that is usually disregarded in the document retrieval. However, the usage emotion is not trivial and requires representation of additional information like stimulus context and eliciting physiology which are necessary for a complete understanding of stimulus' impact. Also, stimuli are delivered to subjects in a sequence through one or more elicitation sessions. These queues or sequences may include temporally overlapping stimuli of different modalities (e.g. visual and auditory) with possible interludes and arranged in a specific order determined by an overseeing expert. Therefore, formal description methods also have to include representation of dynamic stimuli, i.e. time series in one or more multime-dia formats with pauses or breaks between consecutive sequences within the same elicitation session. Since stimuli are used in health related research ethical concerns are of the paramount importance and delivery of improper semantic and emotion content to subjects should be averted. Deficient or incomplete stimuli description is a significant potential cause of elicitation defects because it may lead to retrieval and delivery of an unwanted content. With regard to different modalities, methods for description of other multimedia formats like sound or text can be extended from those applied to visual stimuli. In essence demands on content representation remain the same – high-level semantics with the expected eliciting emotion and the related states should be in the focus of all stimuli description formalisms.

Therefore, the ontology-based paradigm for stimuli annotation and retrieval proposed in the paper consists of terminological and assertional knowledge about high-level multimedia stimuli content and a reasoning engine. These two types of knowledge are the basic components of a knowledge-based system based on Description Logics (DLs) [33] as a set of structured knowledge-representation formalisms with decidable-reasoning algorithms. A variety of tools for knowledge engineering exist [34] which allow construction, management, reuse and reasoning with OWL-based ontologies.

The knowledge base for ontological representation of stimuli has two main components as in the figure below. The terminological component (TBox) describes the relevant notions of the application domain by stating properties of concepts and roles and their interrelations. TBox contains an ontological representation of the knowledge in audio-visual stimuli content. The assertional component (ABox) is a formal set of assertions describing specific semantics or emotion in terms of the terminological knowledge. ABox describes a concrete world by stating individuals and their specific properties and interrelations.

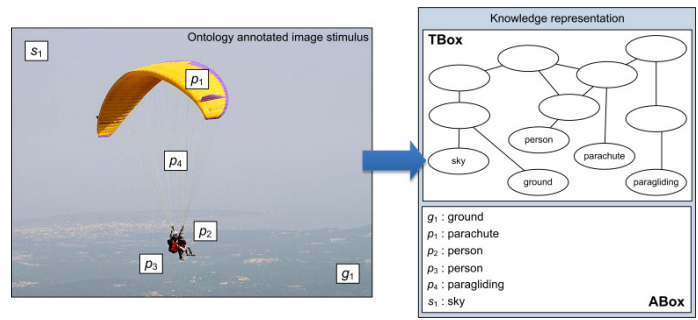

**Fig. 2.** An example of an ontological representation of image stimulus high-level semantics in IAPS picture 8163.jpg (*valence*=7.14, *arousal*=6.53, *keyword*="Parachute"). The extracted information from image analysis constitutes the assertional component (ABox) of the knowledge base and the terminological component (TBox) is defined by the foundation ontology.

The annotation process of a multimedia stimulus begins with identification of concepts in its content that can be observed by subjects and deemed important (by experts) for assessment of stimulus meaning. After a concept is recognized an equivalent concept must also be identified in the ontology used for stimulus representation.

TBox must define all concepts that exist in the stimulus content. After an equivalent concept has been found a new individual is created, associated with the stimulus and stored in ABox. This process is repeated for all stimuli in consideration.

Multimedia content recognition may be performed automatically by an intelligent image recognition algorithm or manually by a group of experts. From the aspect of the STIMONT the image recognition is viewed upon as a black box that yields ground truth object labels. It is of lesser importance how annotations are generated as long as they represent true object labels and relationships present in a multimedia stimulus. However, because of the ethical concerns it would seem highly advisable that an independent team of experts unconditionally verifies quality of stimuli annotation process.

## 3    The STIMONT model

The STIMONT is an upper core ontology designed to provide an integrated and formal description of emotion, high-level semantics, context and physiology content of a multimedia stimulus. The most important feature of the STIMONT is that it provides a formal framework for supporting explicit, human and machine-processable definition of affective multimedia content. This ontology also facilitates storage of stimuli in emotionally-annotated databases, stimuli querying and retrieval and construction of stimuli sequences. The STIMONT enables a common understanding about the perceived meaning of multimedia stimuli, their affective dimensions and context. Using an appropriate interference engine this knowledge allows to derive new facts about the stimuli indexed in affective multimedia databases.

The role of the proposed STIMONT has four aspects:
1. To enable development of intelligent tools for multimedia stimuli selection and retrieval, including improving structure and usability of affective multimedia databases;
2. To serve as a semantic glue or a bridge allowing the integration, i.e. by providing common attachment points, of the different ontologies employed for representation of multimedia stimuli relevant for their application;
3. To serve as a starting point for engineering of other ontologies for description of affectively annotated media, and
4. To provide a reference point for evaluation different ontological approaches in these areas.

The STIMONT is formalism for representation of all relevant knowledge about any type of multimedia stimuli such as images, sounds, video and text. This is achieved by identifying the key aspects of multimedia documents used for stimulation of emotion responses and separately formalizing them as OWL constructs: high-level semantics interpretation, induced emotion and related emotion states, document metadata and emotion-related physiology. This is illustrated in the figure below.

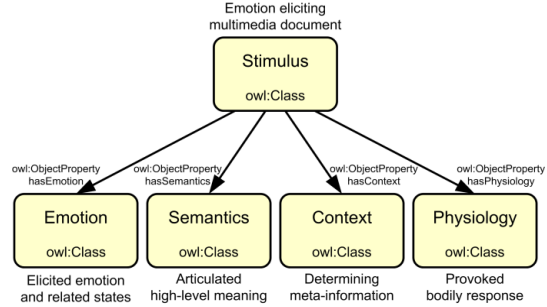

**Fig. 3.** The STIMONT knowledge model defines Stimulus as the top class that represents any multimedia stimulus. Class Stimulus is described with four slots hasSemantics, hasEmotion, hasContext and hasPhysiology for each separate knowledge domain that defines stimulus meaning, eliciting emotion with the related states, information about the document itself and the elicited physiology, respectively.

### 3.1 STIMONT concepts

The core ontology concept is Stimulus which is specified with four other concepts: Emotion, Semantics, Context and Physiology. Each of these specifying components defines one independent information domain that represents particular knowledge about content of a multimedia stimulus. This reflects previously explained approach were a multimedia stimulus elicits some emotion or emotion related states, has a high-level semantics which is articulated by a subject exposed to a stimulus, contains implicit meta-information relevant for representation and provokes a specific physiological response in exposed subject.

Each multimedia stimulus is represented in the ABox as exactly one instance of Stimulus concept. Similarly, Emotion, Semantics, Context and Physiology classes must also have individuals in ABox which are associated with individuals of Stimulus class:

$$Stimulus \equiv \exists_{\geq 1}(Semantics \sqcap Emotion \sqcap Context \sqcap Physiology)$$

Therefore, for a given stimulus at least one of four components must exist to consider the stimulus annotated. Preferably all concepts should be filled for an optimum description because more available information will facilitate multimedia retrieval and construction of powerful and personalized stimuli sequences, which will support a more efficient emotion elicitation process.

The associations of Stimulus class are defined as OWL object properties $hasEmotion.Emotion \sqsubseteq Stimulus$, $hasSemantics.Semantics \sqsubseteq Stimulus$, $hasContext.Context \sqsubseteq Stimulus$ and $hasPhysiology.Physiology \sqsubseteq Stimulus$. Additionally, the ontology differentiates between three types of high-level semantics $Object \sqsubseteq Semantics$, $Scene \sqsubseteq Semantics$ and $Event \sqsubseteq Semantics$. With the STIMONT it is possible to express aggregate semantics with any combination of these three concepts. Furthermore, individuals of Object, Scene and Event can be explicitly retrieved with relations $hasObject \sqsubseteq hasSemantics$, $hasScene \sqsubseteq hasSemantics$ and $hasEvent \sqsubseteq hasSemantics$. The STIMONT supports stimuli retrieval based on objects, scenes and events, emotions, context and physiology objects. Schema of the most important concepts and connections in the STIMONT is given in the figure below.

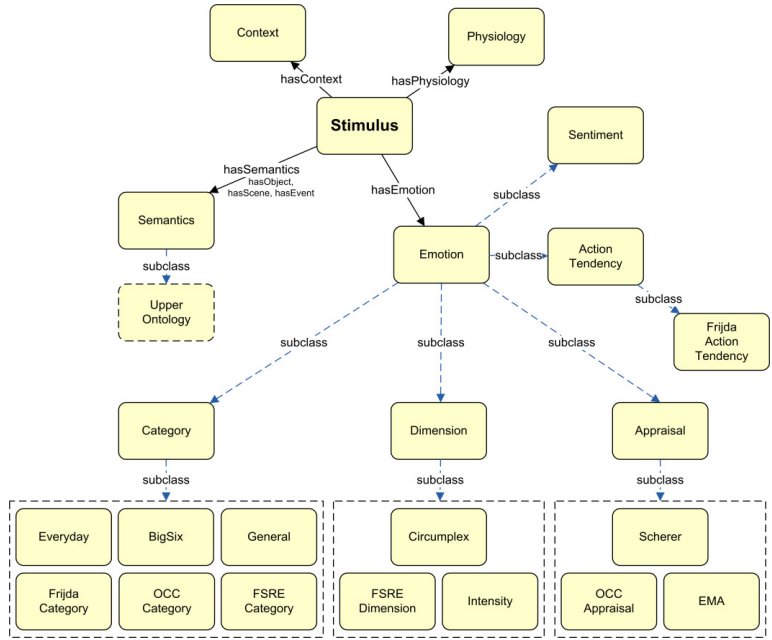

**Fig. 4.** The top ontological concepts and relations in the STIMONT model. Object properties (owl:ObjectProperty) are denoted with solid black lines and class inheritance (rdfs:subClassOf) with dashed blue lines. Semantics can be expressed with any sufficiently expressive foundation ontology.

As can be seen in Figure 4 the STIMONT reuses a singular upper ontology or a combination of upper and domain ontologies for formal representation of aggregate stimuli semantics. This is done intentionally because a number of amicable upper ontologies already exist. Each of them is highly elaborate with several thousands of concepts. It would be wrong, and indeed contrary to the most important ontology goals (such as knowledge reuse), to develop new common-sense ontology just for multimedia stimuli representation. The STIMONT is currently using Suggested Merged Upper Ontology (SUMO) [35] in OWL DL format as foundation ontology for formal representation of stimuli high-level content. SUMO was chosen because of its comparative advantages as the largest freely available common-sense ontology. Other top-level ontologies could also be used for representation of audio-visual content in stimuli semantics if they satisfy criterions of necessary and sufficient expressiveness and decidability. Furthermore, every domain ontology used in the framework has to be aligned to the chosen core ontology thereby ensuring interoperability between different domain ontologies possibly used by the reasoning engine and the retrieval module. But regardless of the actual choice all ontologies must be subclassed to Semantics concept. This preserves the structure of the STIMONT and ensures portability of its stimuli annotations.

### 3.2 Modeling stimuli emotion

The entire emotion taxonomy is in the STIMONT subsumed under the umbrella concept Emotion. For the sake of simplicity and easier understanding the term "affect" was intentionally avoided. Typically, affect is a more general concept than emotion and encompasses a wide range of emotion and related states like feelings, moods, sentiments and attitudes. The term "emotion" is used for description of any type of emotion phenomena: strong and weak, focused and vague, long and short-term. Therefore, in the STIMONT model terms "emotion" and "emotion state" can be used interchangeably with "affect" and "affective state", respectively.

Second-level emotion concepts are Category, Dimension, Appraisal, Action Tendency and Sentiment which constitute domains of categorical and dimensional emotion models, theories of appraisal, action tendencies and sentiment analysis, respectively. Category concepts are further divided into six classes: EveryDay, BigSix, FrijdaCategory, OCCCategory, FSRECategory and General. Each of first five classes formally represents one of the category vocabularies in EmotionML [28]: emotions that frequently occur in everyday life, six primary emotions universal in all cultures, categories related to Frijda's proposal of action tendencies, categories compromising Ortony, Clore and Collins appraisal model, and finally categories used in a study by Fontaine, Scherer, Roesch and Ellsworth, respectively. To provide an even richer expressivity than the EveryDay set, class General was implemented with 66 commonplace emotion concepts like "Euphoria", "Gratitude", "Horror", "Jealousy", "Wonder", etc.

In the stimulus retrieval process, to achieve higher accuracy and recall, it is necessary to assess all emotions assigned to a stimulus. This is accomplished by using roles hasCategory, hasDimension, hasAppraisal, hasActionTendency and hasSentiment which associate Stimulus and classes subsumed by Emotion. Inverse functions are isCategoryOf, isDimensionOf, isAppraisalOf, isActionTendencyOf and isSentimentOf, respectively. Numerical values of emotional dimensions valence, arousal and dominance are very important because they can be found in many stimuli databases annotated with the Russells' Circumplex model such as IAPS, IADS, ANEW, ANET and NAPS. In the STIMONT these values are represented as float datatype properties with domain Dimension and range Stimulus: $valence.Stimulus \sqsubseteq Dimension$, $arousal.Stimulus \sqsubseteq Dimension$ and $dominance.Stimulus \sqsubseteq Dimension$. Float datatatype properties potency, unpredictability and intensity for other dimensional emotion theories are defined in the ontology. Also, datatype properties arousalSD, valenceSD and dominanceSD represent standard deviations for arousal, valence and dominance.

Simultaneous annotation of a single stimulus with different category and dimensional models is permissible in the STIMONT. For example, if a stimulus $stim_1$ from Fig. 2 is annotated with a categorical emotion *Happines* from the BigSix discrete emotion model and also with dimensional values of the PED emotion model (*valence*, *arousal*) = (7.14, 6.53), than the STIMONT's TBox and ABox contain the following axioms:

$$
\begin{aligned}
Category &\sqsubseteq Emotion \\
Dimensional &\sqsubseteq Emotion \\
BigSix &\sqsubseteq Category \\
HappinesBigSix &\sqsubseteq BigSix \\
hasCategory &\equiv \exists Stimulus \sqcap \exists Category \\
hasDimensional &\equiv \exists Stimulus \sqcap \exists Dimensional \\
stim_1 &: Stimulus \\
emo_1 &: HappinesBigSix \\
emo_2 &: Dimension \\
(emo_2, valence) &: = 7.14 \\
(emo_2, arousal) &: = 6.53 \\
(stim_1, emo_1) &: hasCategory \\
(stim_1, emo_2) &: hasDimension
\end{aligned}
$$

**Fig. 5.** The STIMONT's TBox and ABox representing a stimulus *stim₁* from Fig. 2 annotated with an emotion Happines from the categorical BigSix emotion vocabulary (*HappinesBigSix*) and (*valence*, *arousal*) = (7.14, 6.53) values from the dimensional PED model.

To express expert confidence in stimuli potential of eliciting a specific emotion state a string property confidenceLevel with enumerated 5-point Likert scale and a numerical datatype property confidenceValue were implemented in the ontology. Property confidenceLevel indicates the most probable discrete level of agreement in any subsumed class under Emotion with which a particular stimulus instance is connected. The confidence scale ratings are defined using OWL DL and OWL Full construct owl:oneOf. The allowed values under confidenceLevel are string literals VeryHigh, High, Average, Low and VeryLow, where the first represents the highest and the latter the least possible confidence. If a stimulus instance $stim_1$ from the previous example elicits emotion $emo_1$ with an average and $emo_2$ with a very high confidence then the STIMONT's ABox should be extended with the following axioms:

$$
\begin{aligned}
(emo_1, confidenceLevel) &: = "Average" \\
(emo_2, confidenceLevel) &: = "VeryHigh"
\end{aligned}
$$

**Fig. 6.** The stimulus $stim_1$ from the previous example elicits two different emotions *emo₁* and *emo₂* with confidence levels *Average* and *VeryHigh*, respectively.

The second implemented construct for expression of confidence with a real number in the closed interval [0, 1]. This feature is similar to EmotionML attribute "confidence". The STIMONT supports the same expressivity with the datatype property confidenceValue. If confidence in an annotation is unknown or undefined then ABox does not contain confidenceLevel or confidenceValue axioms.

### 3.3 Representation of stimuli semantics

For formal representation of complex stimuli semantics the STIMONT framework relies on Suggested Upper Merged Ontology (SUMO) which was explicitly designed as an upper, core and common-sense ontology. SUMO is currently the largest freely accessible formal upper ontology. Its large knowledge base contains over 25,000 terms and 80,000 axioms. Available SUMO to WordNet mappings help to express concepts in natural language terms [36] which facilitates extension of the framework towards existing tools for informal representation of multimedia (particularly images) with semantic networks and lexical ontologies. As will be discussed further, SUMO possess comparative advantages over other candidate upper ontologies even those who were specifically constructed for formal representation of multimedia. For all afore-mentioned reasons SUMO constitutes particularly adequate choice for multimedia stimuli annotation.

As mentioned previously, to achieve a more detail representation the ontology differentiates between three types of high-level semantics: 1) objects, 2) events, and 3) scenes. The segmentation of high-level semantic components reflects research trends in computer science within areas of concept based image retrieval, automated scene classification and event recognition. The partition of aggregate semantic components should be regarded as mandatory. It helps to create a more explicit and semantically rich representation of multimedia content that leads to more efficient stimuli retrieval. For the purpose of the STIMONT objects can be defined as physical entities represented by stimuli with well-defined and distinguishing features. As such objects may be items or named items such as persons, objects, animals or plants. Events are semantically meaningful human activities (e.g. talking, running, walking, driving, kayaking etc.), taking place within a specific environment and containing a number of necessary objects. Scenes are compound entities which are jointly and implicitly depicted with objects and events in a stimulus.

Typically, an image stimulus will contain a number of physical objects, one scene and at least one event. Based on the current distribution of keywords in affective multimedia databases Object individuals will be the most represented semantic category and the most often used semantics class in search queries. Retrieving multimedia assets in the proposed architecture can be achieved by using semantic query languages such as the SPARQL query language [37]. The next figure illustrates a SPARQL 1.0 query that might be posed by an expert system using STIMONT.

```
PREFIX stimont: http://www.owl-ontologies.com/
stimont.owl#
PREFIX sumo: <http://www.owl-ontologies.com/ sumo.owl#>
SELECT ?stim ?emo ?val ?ar
WHERE
{
  ?stim rdf:type stimont:Stimulus.
  ?stim stimont:hasSemantics ?sem.
  ?sem rdf:subClassOf sumo:GroupOfPeople.
  ?stim stimont:hasDimension ?emo.
```

```
    ?emo stimont:valence ?val.
    ?emo stimont:arousal ?ar.
    FILTER(?val>=6.5 && ?val<=9)
    FILTER(?ar>=1 && ?ar<=3.5)
}
```

**Fig. 7.** A SPARQL 1.0 query for retrieval of positive and calming stimuli ($valence \in [6.5,9] \sqcap arousal \in [1,3.5]$) representing groups of people (sumo:GroupOfPeople).

In the example above stimuli in the lower right corner of the valence-arousal 2D dimensional emotion space defined as $valence \in [6.5,9]$, $arousal \in [1,3.5]$ and with semantics containing some type of groups of people are retrieved from the knowledge base. The query may be executed in the Protégé ontology editor extended with the Jess rule engine [38]. Forward chaining search strategy should be used to maximize the number of returned tuples and the associated multimedia documents.

In another example, IADS stimulus 311.wav is described with a single keyword "Crowd2". Looking at description of concepts in SUMO[1] the most appropriate (i.e. semantically similar) class for this keyword is "GroupOfPeople" ("Any Group whose members are exclusively Humans."). The selected class needs to be in TBox, while ABox contains its instance and an instance of Stimulus concept for representation of the IADS sound file. The Stimulus instance and SUMO's concept GroupOfPeople instance are connected with hasObject property. Therefore, TBox and ABox for representation of IADS stimulus 311.wav would be:

$$hasObject \equiv \exists Stimulus \sqcap \exists Semantics$$
$$stim_2 : Stimulus$$
$$sem_1 : GroupOfPeople$$
$$(stim_2, sem_1) : hasObject$$

**Fig. 8.** A part of the STIMONT's TBox and ABox representing high-level semantics of an IADS stimulus 311.wav that contains sounds of a crowd of people at a sports event. IADS keyword "Crowd2" is translated to SUMO concept "GroupOfPeople" and associated with stimulus instance $stim_2$ as an object instance $stim_1$.

By reusing definitions in the previous example semantically complex multimedia like IAPS pictures 5635.jpg and 7039.jpg in Fig. 1, represented as $stim_3$ and $stim_4$ in ABox, can be annotated with several different concept individuals as:

$$stim_3, stim_4 : Stimulus$$
$$sem_2 : WinterSeason$$
$$sem_3 : Snow$$
$$sem_4 : Street$$
$$sem_5 : City$$
$$sem_6 : Automobile$$
$$sem_7 : Covering$$
$$sem_8 : TransportationDevice$$

---

[1] http://www.ontologyportal.org/SUMO.owl

$$sem_9 : GeographicArea$$
$$sem_{10} : FallSeason$$
$$sem_{11} : Forest$$

**Fig. 9.** Representation of complex high-level semantics in IAPS pictures 5635.jpg and 7039.jpg from Fig. 1. The original keywords describing scenes "WinterStreet" and "Train" were expanded to include objects visible in the pictures and considered important for cognitive processing of the content and emotion stimulation. Semantics individuals $sem_2 – sem_7$ are connected with *hasObject* property to Stimulus individuals $stim_3$ and $sem_8 – sem_{11}$ to $stim_4$ which is not shown in this example.

As mentioned previously, contemporary affective multimedia databases lack expressivity in description of stimuli. Pictures 5635.jpg and 7039.jpg are annotated in IAPS with their respective keywords "WinterStreet" and "Train" that convey only scenes in global terms while individual objects in the pictures are neglected. With STIMONT it is possible to assign an arbitrary number of descriptors for each stimulus. However, there is no uniform standard which determines the optimal level of expressivity. The required level of annotation expressiveness is determined by experts and database owners. It may include only one semantically predominant concept (as in IAPS, IADS, GAPED, ANEW etc.), or a concept and its subsuming category (as in NAPS), or – ultimately – the description may include all scenes, objects, events and even affective terms that contribute to meaning of a picture. The example in Fig. 9 follows the last suggestion and describes IAPS pictures 5635.jpg and 7039.jpg with SUMO concepts "WinterSeason", "Snow", "Street", "City", "Automobile", "Covering" and "TransportationDevice", "GeographicArea", "FallSeason", "Forest", respectively. If some upper ontology other than SUMO is used for high-level annotation, the concept corpora would be different.

Pictures without a semantic context are much simpler to define than pictures with complex meaning. Retrieval of documents with narrow context is less prone to errors since their meaning is less noisy and the semantics can be described more accurately. IAPS picture 8163.jpg in Fig. 2 is a clear example of such a document. This picture is originally described with only one IAPS keyword "Parachute", but in Fig. 2 this is expanded to 5 different unsupervised keywords "Ground", "Parachute", "Person", "Paragliding", "Sky" that were translated to subsuming SUMO concepts "LandArea", "Device", "Human", "Transportation", "AtmosphericRegion" as in the next figure. The description requires 5 concepts in TBox and 6 individuals in ABox since the picture displays two persons, i.e. instances of "Human" concept.

$$sem_{12} : LandArea$$
$$sem_{13} : Device$$
$$sem_{14} : Human$$
$$sem_{15} : Transportation$$
$$sem_{16} : AtmosphericRegion$$

**Fig. 10.** A part of the STIMONT's TBox and ABox representing high-level content of narrow semantic context IAPS stimulus 8163.jpg. All Semantics individuals $sem_{12} - sem_{16}$ are connected to $stim_1$ with *hasObject* property. Together with ontological annotations in Fig. 2 this example comprehensively describes semantics and emotion of this stimulus.

### 3.4 Stimuli context and elicited physiology

Emotion and semantics are not enough to reach a complete understating of a multimedia stimulus. Knowledge about emotion and semantic can be derived by observation, i.e. empirically, of a multimedia document. Additional information about the expected impact of stimulus must be stated by other sources. This metadata is represented by the concept Context in the STIMONT.

Therefore, Context class can be regarded as a container for storage of all data relevant for understanding of a stimulus that does not belong to Semantics or Emotion classes. Using STIMONT it is possible to annotate stimuli with a wide range of additional metadata such as: stimulus unique identifier (id), stimuli database name, document dimensions (i.e. width and height), document size, color depth, stimulus length (in seconds; applicable to dynamic media such as sounds and films), author name, legal owner, creation date and time, depicted location, area or region, stimulus media format etc. Also, name of stimuli database is very important because it enables functional integration of different affective multimedia databases. Context class can be used to express this versatile knowledge domain and ensure a more complete stimuli description.

For example, the IADS stimulus 311.wav from Fig. 8 is 6 seconds long. This meta-information can be expressed with STIMONT to achieve a better understanding of the stimulus. The extension of representation of the stimulus 311.wav using Context class is demonstrated below.

$$
\begin{aligned}
hasContext &\equiv \exists Stimulus \sqcap \exists Context \\
stim_2 &: Stimulus \\
context_1 &: Context \\
(context_1, dbName) &:= \text{"IADS"} \\
(context_1, id) &:= \text{"311"} \\
(context_1, length) &:= 6 \\
(stim_2, context_1) &: hasContext
\end{aligned}
$$

**Fig. 11.** A segment of the STIMONT's TBox and ABox representing metadata about sound stimulus 311.wav from the IADS database. The stimulus is 6 seconds long. Stimuli metadata is represented with Context class.

Simple level of Dublin Core metadata element set [39] is a standard for cross-domain information resource description and can ensure simple and standardized set of conventions for description of these additional context data. Stimuli metadata can be encoded with Dublin Core and integrated into STIMONT using the following pattern: *dc:type* describes the type of the resource, *dc:creator* refers to a person or body responsible for the content of the resource, *dc:contributor* is a person or entity responsible for making contributions to the content of the resource, *dc:date* is associ-ated with an event in the lifecycle of the resource and finally *dc:format* describes the multimedia format of the resource. However, if Dublin Core is used for representation of stimuli context it must be strongly constrained. Although an OWL version of the Dublin Core is suitable for integration with enterprise ontologies its properties are loosely annotated. For instance, the preferred usage of the *dc:creator* tag is a "LastName, FirstName" literal. This is potentially ambiguous because different persons can have the same name (i.e. John Smith). Within the STIMONT this problem can be overcome if an instance of a Person class from a common upper ontology is used instead of a literal. However, such *ad hoc* solution negates Dublin Core interoperability and actually replaces loose confinement with another limitation. Currently this is an open problem and may be solved in the future by adding all contextual properties directly in STIMONT and completely avoiding domain ontologies.

Time and location metadata are important because stimuli with pronounced personal meaning (i.e. ego relevance) are better at affect elicitation. Stimuli semantics aligned with personal experiences results in a stronger appraisal, intense cognitive associations and generates more pronounced behavioral reaction in exposed subjects. For these reasons time and location metadata can help to reject unimportant or meaningless semantics and retrieve only those stimuli that are spatially and temporarily associated with a subject's anxiety, phobia or mental trauma. For example, in a stimulation of a victim of the September 11 terrorist attack notions like "Location = New York, USA" and "Time = September 11, 2001" will confine the search to the most relevant stimuli.

If the SUMO is used as the upper ontology then the notion of time is represented with the SUMO concept "TimeMeasure" and its subsumed classes, while geographic location can be represented with suitable domain ontology. FAO Ontology and Terminology System[2] is a geopolitical ontology written in OWL DL that represents a hierarchy of the world's countries. This ontology also includes a multitude of useful information about geopolitical concepts like international names, nationality, latitude, longitude, list of countries bordering states and also properties "isPredecessorOf" and "validUntil" that together define political history of a concept.

Definition of Context class should not be regarded as static. If needed new metadata important for emotion elicitation may be added to this container concept. Context is important for complete description of pictures, but critical for correct understanding of pictures as stimuli.

Since affective multimedia stimulates generation of physiology in exposed subjects and physiology has been shown to be an important and objective channel for automated estimation of emotion and related states [40], it is useful to connect physiology data with Stimulus class. In a typical emotion stimulation experiment physiology channels such as heart and breathing rate, skin temperature and conductance, ECG and EEG are acquired with a suite of sensors and the data is stored in formatted files. The files are usually large (tens or hundreds of megabytes), have diversified structures and contain series of numerical values recorded at a specific sampling rate. Ontology is not an appropriate medium for direct representation of this type of data, but ontological constructs can be successfully used to reference physiology files stored at some suitable and accessible location.

---

[2] http://www.fao.org/docrep/008/af243e/af243e00.htm

In the STIMONT Physiology concept has only one parameter path that references an URI with a physiology data file acquired during emotion elicitation experiment. Attribute path is defined as owl:DatatypeProperty. It is associated with Stimulus with domain constraints and additionally with owl:ObjectProperty hasPhysiology with domain Stimulus and range Physiology ($hasPhysiology.Physiology \sqsubseteq Stimulus$). The attribute hasPhysiology is non-functional since one stimulus can provoke an unrestricted number of physiology responses ($\bot \sqsubseteq\, \leq 1 hasPhysiology$). The STIMONT does not make any restrictions to the format of physiology files, acquisition procedures or methods for their distribution. In a typical scenario a set of physiology channels would be recorded in a psychophysiology laboratory environment using acquisition equipment and data loggers. Afterwards the data can be processed and a set of acquired physiology objects with annotations attached to each stimulus object can be made available online. It is assumed that each resource has unique URI.

For example, ABox with acquired physiology for two subjects that were exposed to the IAPS picture 8163.jpg in Fig. 2 is shown below. Three distinct physiologies were acquired ($phy_1$, $phy_2$, $phy_3$). Heart rate (HR) channel was recorded for both subjects and skin resistance (SR) only for the first one. Subjects were stimulated using the same image stimulus ($stim_1$). The image's unique identifier (id) is "8163".

$$
\begin{aligned}
stim_1 &: Stimulus \\
phy_1 &: Physiology \\
phy_2 &: Physiology \\
phy_3 &: Physiology \\
context_1 &: Context \\
(context_1, id) &: = \text{"8163"} \\
(context_1, dbName) &: = \text{"IAPS"} \\
(phy_1, path) &: =\text{"http://www.foo.com/subject1\_hr"} \\
(phy_2, path) &: =\text{"http://www.foo.com/subject1\_sr"} \\
(phy_3, path) &: =\text{"http://www.foo.com/subject2\_hr"} \\
(stim_3, phy_1) &: hasPhysiology \\
(stim_3, phy_2) &: hasPhysiology \\
(stim_1, context_1) &: hasContext
\end{aligned}
$$

**Fig. 12.** An example of ABox representing the annotation of three distinct physiologies (heart rate and skin resistance) acquired from two subjects stimulated with the same IAPS image 8163.jpg.

## 4 Ontology implementation and validation

The STIMONT was written in OWL DL using the Protégé ontology editor (version 3.4.5). Currently the ontology has 178 classes and 38 properties. By far the most complex part of the ontology is related to the Emotion class with 173 subsumed classes which represent individual emotion theories and related concepts such as ActionTendency, Appraisal, Category, Dimension and Sentiment. Representations of Appraisal, Dimension and Sentiment are expressed as float values in range [0.0, 1.0] and defined in domain datatype properties, while ActionTendency and Category are represented as individuals in ABox. Object properties like hasActionTendency, hasAppraisal, hasContext, hasEmotion, hasPhysiology and hasSemantics connect individuals of Stimulus class to individuals of pertaining classes.

To facilitate intuitive retrieval of affective multimedia documents, a software tool Intelligent Stimuli Generator (intStimGen) was developed. The intStimGen was designed as n-layered Desktop application with an efficient user-friendly graphical interface.

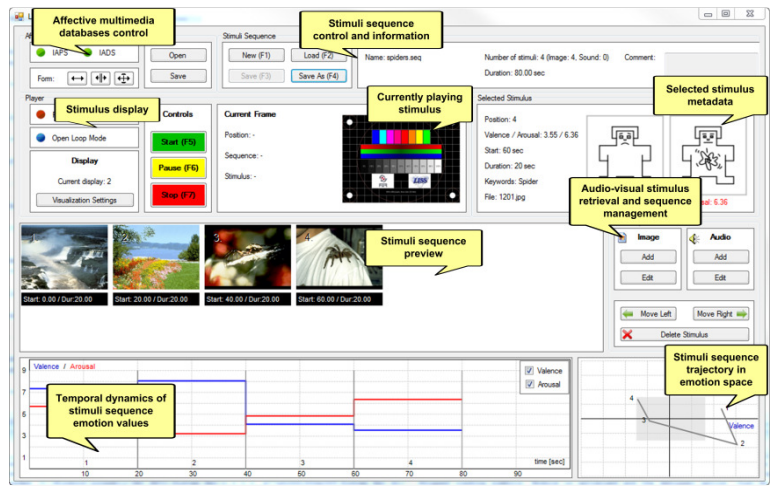

**Fig. 13.** The main screen of the Intelligent Stimuli Generator tool displaying a fear-provoking sequence constructed with IAPS pictures. The pictures were retrieved by the tool based on semantics related to arachnophobia.

This tool enables searching of affective multimedia databases and construction of stimuli sequences using semantic and emotion descriptors. It follows previous work in developing an online tool WNtags for collaborative annotation and retrieval of IAPS images using WordNet [41]. In intStimGen stimuli semantics can be described with unrestricted keywords, tag clouds, WordNet synsets and SUMO concepts while emotion descriptors are based on the dimensional model. Relatedness between descriptors can be calculated with lexical and semantic similarity measures. The implemented measures are: Levenshtein, path length, Wu Palmer, Leacock Chodorow and Li algorithm [42]. Other measures can be modularly added later if needed. Document retrieval and ranking can be performed using any combination of descriptors and similarity measures. Integrated exploration of IAPS and IADS is supported, and the application's architecture enables other stimuli databases to be modularly added in the future. The constructed sequences can be displayed to a participant on a separate screen and his physiological and behavioral responses may be acquired using specialized hardware. During exposure the application sends system messages which can be used for synchronization of physiology acquisition equipment and other systems.

The tool is written in .NET 3.5 framework and optionally may use Jena.NET toolkit[3] for querying and integration with the Protégé. In this environment retrieval can be performed with the Pellet 2.2.2 reasoner over XML-based DIG 1.1 query interface. SPARQL queries as in Fig. 7 may be entered using Jess system shell.

The STIMONT is distributed as a single OWL DL file "stimont.owl" and is available for research purposes together with stimuli metadata used in the experiment and installation of the intStimGen tool [43].

**4.1 Multimedia stimuli retrieval**

In order to functionally validate the proposed concept it was necessary to evaluate the retrieval performance of STIMONT and compare it to other methods for annotation of multimedia stimuli. The retrieval process from the existing affective multimedia databases is based entirely on unmanaged keywords and affective values. Additional metadata like context and physiology are not supported. Therefore, to objectively compare the ontology-based retrieval with the existing procedures it is necessary to confine the validation on the shared search modalities, i.e. semantics and emotion. Furthermore, since emotions in affective multimedia databases are described as vectors in Euclidian space [18, 19] with well known distribution while high-level multimedia meaning is more uncertain, semantics is a more important parameter in evaluation.

For experimentation in multimedia retrieval a subset of 772 pictures were first extracted from IAPS. The selected pictures have clear meaning and do not provoke intense negative emotions. In the next step high-level content of the selected pictures was hand annotated with WordNet terms. The keywords were extended with the most appropriate WordNet synsets by considering content of each picture and the meaning of its keyword. If an identical synset does not exist for a given keyword it was extended with more than one synset. Then WordNet to SUMO mappings [44] were used to obtain semantically equivalent SUMO annotations.

For this experiment 500 pictures were selected that show animals, nature, people, household objects and food. Each picture in document collection **D** was annotated with an original IAPS keyword $w_i \in \mathbf{W}$ and the transformed SUMO concept $c_i \in \mathbf{C}$. In total, the selected pictures were described with 318 individual keywords and 162 SUMO concepts. The selected images are emotionally either neutral or highly positive and arousing. Picture affective values were inherited from IAPS. The query collection **Q** consisted of different individual queries from both annotation glossaries. The retrieved pictures were ranked using string inclusion, Levenshtein, path length and Wu Palmer algorithms. For each query $q_i \in \mathbf{Q}$ a subset of documents $\dot{\mathbf{D}} \subset \mathbf{D}$ with $|\dot{\mathbf{D}}| = 100$ pictures were randomly selected and classified using the aforementioned glossaries and relatedness measures. In total 120 such document retrieval tasks were executed and 3000 pictures were ranked and classified. The queries always contained one keyword or ontology concept. Only terms that exist in the annotations of $\dot{\mathbf{D}}$ were used thereby ensuring that the retrieved set should be nonempty.

---

[3] http://www.linkeddatatools.com/downloads/jena-net

Each of the ranking algorithms provided a measure of relatedness (similarity score) between two labels rel(x, y) ∈ [0,1], where x, y = **W** ∪ **C**, that was used to rank the retrieved images, with properties:

$$\text{rel}(x, y) = 1, x = y$$
$$\text{rel}(x, y) < 1, x \neq y$$

Since unsupervised keywords are mutually semantically unrelated only lexical similarity algorithms (string inclusion and Levenshtein) were used to establish a relatedness measure between them. Path length and Wu Palmer algorithms were used exclusively on SUMO knowledge taxonomy with ontology concepts as nodes and concept inheritance (i.e. IS-A functional relationship) representing connections between nodes. The inclusion represents the simplest measure that only checks if a string is included or exists in another string, while Wu Palmer is the most complex. The idea behind this choice of algorithms was to test how more simple description methods such as keywords and lexical similarity measures compare with more complex annotations (i.e. ontologies) and compatible retrieval methods.

Since each query $q_i$ always returned all $|\dot{D}| = 100$ images (there was no cut-off) in descending order, with picture most relevant to the search query on top of the list and the least relevant at the end, it was necessary to determine the threshold value $t$ that determines the classification boundary and how the retrieved pictures will be categorized. Given a retrieved picture $p_i \in \dot{D}$ and its rank in the set of retrieved pictures $r_i \in [1, 100]$, where $r_j = 1$ indicates the most relevant picture $p_j$, the picture $p_i$ was classified in category $cat_i = \{True, False\}$ as:

$$cat_i = \text{True}, r_i \leq t$$
$$cat_i = \text{False}, r_i > t$$

Therefore, in each query the system partitioned the retrieved corpus into two subsets of pictures: those it considered relevant to the search query (category True), and those it did not (category False). The threshold value was determined for each query using the lift curve. For the maximum precision in retrieval the threshold was set to the rank with the highest lift factor. This adaptive approach enables more objective evaluation of different ranking retrieval algorithms than a constant classification threshold. For example, the result set 1 in [43] containing pictures queried with SUMO concept "Man" and ranked with Wu Palmer algorithm has the maximum lift factor for $r = 5$ implying that to achieve the highest precision in this set only pictures with $r \leq 5$ have to be classified as True and all other pictures with $r > 5$ as False. The order in which the returned documents were presented was disregarded because it was not possible to exactly establish the optimal order of documents, but only the partition of documents into the two classes: relevant (True) and non relevant (False). The aggregated experiment results are shown in the next table.

**Table 1.** Aggregated experiment results with measures of information ranked retrieval quality [43].

|  |  | **Accuracy** | **Precision** | **Recall** | **Fall-out** | **F-Measure** |
| --- | --- | --- | --- | --- | --- | --- |
| Keywords | Inclusion | 0.7580 | 0.5887 | 0.3279 | 0.6721 | 0.6103 |
|  | Levenshtein | 0.7990 | 0.5936 | 0.2555 | 0.7445 | 0.6501 |
| Ontology | Path length | 0.8567 | 0.8381 | 0.3752 | 0.6248 | 0.8301 |
|  | Wu Palmer | 0.8436 | 0.8123 | 0.3863 | 0.6137 | 0.8106 |

As can be seen in the table, ontology annotations are better than keywords in all aspects of ranked retrieval of IAPS pictures, albeit the difference is small in some parameters. Accuracy is 4.4557 – 9.8657% higher and precision even more 21.872 – 24.9369%. Recall is 4.7381 – 13.0802% better and F-measure 16.0447 – 21.9814%. Since Fall-out is the proportion of non-relevant documents that are retrieved, out of all non-relevant documents, lower result is better. F-measure combines precision and recall as their weighed harmonic mean. Taken together the results could be interpreted that the ontology retrieval is superior in discrimination of non-relevant documents while a bit less capable in identification of relevant documents.

The highest gain with ontology-based retrieval is in precision (24.9369%) and F-measure (21.9814%) because it benefits from knowing the semantic relatedness between different concepts which is beyond the capabilities of lexical algorithms. A single concept may encapsulate meaning of several different keywords which must be individually identified and used in retrieval.

String inclusion algorithm has the worst aggregated performance and represents the baseline algorithm that can be used on contemporary semantic descriptions in affective multimedia databases. However, it has high true positive rate, resulting in above average accuracy and recall, only if query keyword is used for annotation of semantically identical images. This algorithm requires that user has some conception about the keywords before using the database otherwise the retrieval performance will be very poor. All other algorithms are an upgrade to affective multimedia databases and can be used only if they are extended with additional retrieval tools as was done in this experiment. In this experiment the Levenshtein algorithm represented the best retrieval method based purely on lexical properties of picture descriptions. Since IAPS keywords are only lexically related more meaningful description of picture semantics should be implemented with ontologies. Using Levenshtein algorithm the difference between keywords and ontology is somewhat reduced, especially in accuracy, but ontology still gives better results, especially in the percentage of relevant documents that are returned. As can be seen in Table 1 there is a smaller difference in retrieval performance between path length and Wu Palmer algorithms as is between ontology and keywords. This indicates that the relatedness algorithm is much less important than the method for annotation of affective pictures.

The results also indicate that IAPS pictures are weakly annotated. Only one keyword or ontology concept per picture is insufficient to relate the full higher meaning which results in higher false positive and false negative count in some data sets. Furthermore, statements about multimedia content – without rules about the domain – are not enough to encapsulate all knowledge important for quality retrieval.

This experiment should be regarded as an initial validation of the ontology presented in the paper. The tested application of the ontology is a minimum that an ontology-based description of affective multimedia may provide. More complex retrieval procedures involving procedural knowledge and more expressive ontological annotations of stimuli content should be evaluated in the future.

## 5 Discussion and future work

In practical terms the STIMONT should be regarded as a method for annotation of emotion eliciting multimedia to enable better and more efficient stimuli knowledge representation and retrieval that lead to the development of tools for improvement of contemporary affective multimedia databases. The STIMONT builds upon the work of W3C in designing the EmotionML and extends it with additional vocabularies and expressivity towards describing multimedia documents with eliciting emotion values. In this regard the STIMONT's multimedia annotations are akin to EmotionML references "triggeredBy" or "targetedAt". The ontology does not differentiate between semantics of these two references. Which specific reference of these two is implied depends weather ABox contains individuals of emotion and appraisals or action tendencies. Semantics of EmotionML references "expressedBy" and "experiencedBy" is not currently supported because the STIMONT is primarily designed for representation of stimuli (i.e. anticipated emotion) and not of emotion experienced by subjects. If in the future the proposed ontology could be expanded to formalize the entire spectrum of emotion-related phenomena then it may also be possible to represent observable behavior expressing the emotion and the subject experiencing the emotion.

In this application the ontology could deduce various relationships, in particular equivalence, between seemingly implacable emotion terms. For example, using axioms it is possible to state that basic emotion anger in the Ekman's BigSix and the OCC model are the same as $anger.BigSix \equiv anger.OCCCategory$. Another emotion theory FSRE with the same emotion category anger may also be considered. If BigSix and FSRE anger emotion categories are identical than it is possible to state $anger.BigSix \equiv anger.FSRECategory$. Then by using a reasoning engine it can be immediately inferred that $anger.FSRECategory \equiv anger.OCCCategory$ is also true. Such equivalencies or other relationships (e.g. subsumption) are possible for other emotion concepts as well. However, cross-equalization of different emotion vocabularies within the same knowledge base should be accomplished by a knowledge engineer working in tandem with a qualified domain expert. Different categorical emotion theories should not be conjoined just because they have lexically similar terms. More thorough analysis of vocabulary semantics is warranted before these identities are axiomatized and stored in TBox. Integration of WordNet-Affect [45] taxonomy of emotion related terms in the STIMONT would be beneficial for establishment of relationships and transformations between different emotion concepts. However, the problem of correct identification of concept domain and range still remains.

As mentioned in Section 3.3 the top stimuli semantics is described with SUMO because of its many useful and practical features. Other upper ontologies like DOLCE [46] and ConceptNet [47] could be used instead of SUMO for representation of high-level objects, events and scenes. Furthermore, dedicated multimedia-representation ontologies like Large-Scale Concept Ontology for Multimedia (LSCOM) [48] and Core Ontology for MultiMedia (COMM) [49] could also be a natural choice for annotation of multimedia stimuli. High accuracy concept-based video retrieval in some closed domains has been reported for these top ontologies [50]. However, compared to SUMO other ontologies have the following disadvantages when used for high-level stimuli annotation:

1. Fewer concepts in TBox (i.e. smaller vocabulary);
2. Lesser expressivity;
3. Proprietary or limited access for ontology reuse requiring individual permissions, and
4. Lack of integration with other knowledge representations used in image annotation frameworks based on semantic networks such as WordNet [51].

The presented ontology captures high-level information in semantics and emotion but it has been shown that low-level features like color, hue, brightness and texture influence perception and can be positively correlated to arousal and discrete emotion categories [52]. Therefore, it future version of the STIMONT it would be useful to represent perceptual and low-level meaning of stimuli. This would certainly add to the richness of representation and enhance multidimensional stimuli retrieval.

With regards to the terminology, throughout the paper term "subject" was used to denote a person who is exposed to an emotionally annotated multimedia document. There was no intention to limit the scope of this term only to participants of an emotion or attention related experiments. The STIMONT can be used in Human-Computer Interaction (HCI) or any other interaction between humans and computer that involves emotion and emotion related phenomena such as affective interfaces or video games. Also, the STIMONT can be applied for representation of multimedia and Virtual Reality (VR) stimuli in psychotherapy as in exposure therapy (ET) or stress inoculation training (SIT) [53]. In this regard terms "subject", "exposed person", "participant" and even "trainee" could be used as synonyms.

The STIMONT should not be seen as a singular or exclusive solution to the problem of formal representation of emotions and related states, but rather as a representative of knowledge representation methods based on DLs specifically intended for construction of practical software tools for affective multimedia databases.

Due to pleura of different emotion models, and the ongoing active research on the origin and nature of emotions, it is technically impossible to create a single ontology that will equally and sufficiently represent all theories. For this reason it is important to allow for future expansion and enlargement of the ontology while retaining backward compatibility with previous versions. Other solutions in this domain are possible, but the STIMONT offers numerous advantages like logic-based reasoning, formal presentation of concepts, inclusion of leading emotion theories, standardized representation vocabularies and hierarchies of affective terms with ready-to-use WordNet mappings.

In the future STIMONT should allow detailed representation of dynamic stimuli i.e. individual frames of videos and movies. Films, video-clips and related dynamic presentation technologies are often used for induction of emotion in the laboratory because of their relatively high attention capture and intensity [54]. Compared to other multimedia formats personalized video clips, in particular real-life footage, represent the most powerful multimedia elicitation tool because they provide a high degree of personal significance and immersion that that results in a more intensive stimulation than with any individual audio or visual stimuli. Therefore, integration of the STIMONT with MPEG-7 multimedia content description standard for affective annotation of video would be beneficial. Finally, besides annotation of affective multimedia, which is the STIMONT's immediate application, this ontology can also be used to facilitate integration of different emotion theories and for general reasoning about emotion-related phenomena. This line of research is primarily oriented towards creating tools in areas of psychology and neuroscience.

## 6 Conclusion

The proposed ontology offers a number of advantages over the existing methods for knowledge representation in the emotionally annotated multimedia. In the evaluation experiment the ontology-based approach achieved up to 24.9369% better precision in ranked retrieval of IAPS pictures. Compared to the existing method of retrieval with keywords this represents an increase in all performance and correctness measures. Although ontologies require more complex prerequisites and reasoning infrastructure than the markup or keyword based annotations, they are much more advantageous than the simpler methods and provide for a better multimedia retrieval.

The STIMONT provides a practical model for formal representation of high-level semantics, emotion and related states, document context and stimulated physiology that collectively define a multimedia stimulus. The presented ontology builds upon W3C EmotionML format and extends it with additional emotion vocabularies. The STIMONT reuses SUMO common sense ontology and SUMO to WordNet mappings to provide a rich high-level semantic expressivity with interface to commonplace models based on informal knowledge and probabilistic reasoning. The STIMONT facilitates knowledge reuse, interoperability and formalization of stimuli information which are superior to the contemporary methods for representation of emotionally annotated documents. All these features enable formal, consistent and systematic annotation of the affective multimedia and DL-based reasoning about their aggregated content and document properties.